1

# A New Method for Obtaining the Baryons Mass under the Killingbeck Plus Isotonic Oscillator Potentials

NASRIN SALEHI*

Department of Basic Sciences, Shahrood Branch, Islamic Azad University, Shahrood, Iran
E-mail*: Salehi@shahroodut.ac.ir

*Abstract*. In this work, the spectrum of ground state and excited baryons (N, $\Delta$, $\Lambda$, $\Sigma$, $\Xi$ and $\Omega$ particles) has been investigated by using a non-relativistic quantum mechanics under the Killingbeck plus isotonic oscillator potentials. Using the Jacobi-coordinates, anzast method and generalized Gürsey Radicati (GR) mass formula the three body wave equation is solved to calculate the different states of the considered baryons. A comparison between our calculations and the available experimental data shows that the position of the Roper resonances of the nucleon, the ground states and the excited multiplets up to three GeV are in general well reproduced. Also one can conclude that; the interaction between the quark constituents of baryon resonances could be described adequately by using the combination of Killingbeck and isotonic oscillator potentials form.

## 1. INTRODOCTION

The hadrons spectroscopy is very important to study its structures and the nature of the interacting forces between its constituents. In quark models, the baryons are three-quark states and there are number of very different model calculations for the baryon masses [1-8]. Such a picture of these elementary particles has been having success in explaining and describing spatial ground state of the flavor *SU*(3) vector mesons and baryon octet. But in recent years, baryon spectroscopy has attracted many interests because baryons were the focal point of quark model development [9, 10]. Such a system can be studied by Quantum Chromo Dynamics which describes what between quarks and gluon and their interactions [11, 12].

In order to study massive baryons there are two options, experimentally and theoretically. For example, experimentally mass spectrum of singly charmed heavy baryons is well known but the others are not. Recently, A. M. Abazov and T. Aaltonen have published articles separately in which there are mass measurements of singly bottom baryon $\Xi_b^-$ by the $D\emptyset$ [13] and CDF [14] collaborations. On the other hand, from lattice QCD point of view, there are interesting efforts about quenched calculations such as what is done by K.C. Bowler et al, in which they presented the results of an exploratory lattice study of heavy baryon spectroscopy [15] or R. Lewis et al calculated Masses of singly and doubly charmed baryons in quenched lattice QCD using an improved action of the D234 type on an anisotropic lattice [16] and or N. Mathur et al. computed the mass spectrum of charmed and bottom baryons on anisotropic lattices using quenched lattice nonrelativistic QCD [17] also S. Gottlieb and S. Tamhankar published results from a lattice study of the semileptonic decay of $\Lambda_b \to \Lambda_c l \nu_l$ [18], A. Ali Khan et al. presented lattice results for the spectrum of mesons containing one heavy quark and of baryons containing one or two heavy quarks. [19] and in this discipline reader can check Ref. [20] and for dynamical sea quark flavor simulations, H. Na and S. Gottlieb studied the heavy baryon mass spectrum on gauge configurations that include 2+1 flavors of dynamical improved staggered quarks [21] and they present results for the mass spectrum of charm and bottom heavy baryons, using MILC coarse lattice configurations with 2+1 flavors [22] and also R. Lewis and

R. M. Woloshyn Bottom calculated baryon masses based on a 2+1 flavor dynamical lattice QCD simulation. Of course that for the heavy baryon mass spectrum and most results is in fair agreement with observed values [23]. On the other hand motivation of studying light baryons is that it enables us to find an understating of the structures and their interactions [24]. Actually to have fundamental manifestation for the long-distance quark and gluon dynamics that is governed in QCD, we use hadron mass spectrums. As matter of fact Non-perturbative calculations and numerical simulation in space-time lattice [25]lead to method to get to this matter from QCD Lagrangianwithout having any approximations, this results to determination of light quark masses as well [26].In the recent years considerable effort have been done in the lattice QCD calculation of the light hadron spectrum[27]. We can mention to a calculation in which mass of hadron has been calculated with accuracy of 0.5%-3% [28].

As a matter of fact, the three Quark interaction can be divided in two parts: the first one, containing the confinement interaction, is spin and flavour independent and it is therefore *SU* (6) invariant, while the second one violates the *SU* (6) symmetry [1, 29-32]. It is well known that the Gürsey Radicati mass formula [33] describes quite well the way *SU* (6) symmetry is broken, at least in the lower part of the baryon spectrum. In this work we applied the generalized Gürsey Radicati (GR) mass formula which is presented by Giannini and et al [29] to calculate the baryon masses. The model we used is a simple Constituent Quark Model in which the *SU* (6) invariant part of the Hamiltonian is the same as in the hypercentral Constituent Quark Model [35, 36] and the *SU* (6) symmetry is broken by a generalized GR mass formula. In this paper the exact solution of the Schrodinger equation for the Killingbeck plus quantum isotonic oscillator potentials [37-39] via wave function ansatz is given and we introduce the generalized GR mass formula, then we give the results obtained by fitting the generalized GR mass formula parameters to the baryon masses and we compare our calculation spectrum with the experimental data.

## 2. THE USED THEORETICAL MODEL

The Hamiltonian of the Schrödinger equation is as the following form

$$H = \frac{-\hbar^2}{2\mu}\nabla^2 + V(r), \qquad (1)$$

Where

$$\nabla^2 = \left[\frac{d^2}{dr^2} + 5r^{-1}\frac{d}{dr} - \frac{\gamma(\gamma+4)}{r^2}\right], \qquad (2)$$

Since the Schrödinger equation is

$$H\psi_{\nu\gamma} = E\psi_{\nu\gamma}, \qquad (3)$$

So in the six-dimensional, these equations for a system containing three quarks with a potential $V(r)$ and by considering of $\psi_{\nu\gamma} = u_{\nu\gamma}(r)r^{-5/2}$ can be written as:

$$\frac{d^2u_{\nu\gamma}(r)}{dr^2} + 2\mu\left[E - V(r) - \frac{(2\gamma+5)(2\gamma+3)}{8\mu r^2}\right]u_{\nu\gamma}(r) = 0, \qquad (4)$$

Where $u_{\nu\gamma}(r)$, $r$ and $\gamma$ are the hyperradial wave function, the hyperradius and the grand angular quantum number, respectively. $\gamma$ is also given by $\gamma = 2n + l_\rho + l_\lambda$, $0 \leq n \leq \infty$ with the angular momenta $l_\rho$ and $l_\lambda$ which are associated with the Jacobi coordinates ($\vec{\rho}$ and $\vec{\lambda}$) [40] and $\nu$ denotes the number of nodes of three quark wave functions. In Eq. (4) $\mu$ is the reduced mass which is defined as $\mu = \dfrac{m_\rho m_\lambda}{m_\rho + m_\lambda}$ in which $m_\rho = \dfrac{m_1 m_2}{m_1 + m_2}$, $m_\lambda = \dfrac{3m_3(m_1 + m_2)}{2(m_1 + m_2 + m_3)}$, $m_1$, $m_2$ and $m_3$ are the constituent quark masses [36]. In our model, the interaction potential is assumed as

$$V(r) = ar^2 + br + \frac{c}{r} + \frac{d}{r^2} + \frac{hr}{r^2+1} + \frac{kr^2}{(r^2+1)^2}, \qquad (5)$$

Cornell interaction (Coulomb plus linear) which is static and spherically symmetric interaction, has a physical application in Mesonic systems ie. Charmonium and Bottomonium. Coulomb-like part potential is a short range potential that arises from exchanging a massless gluon between the quarks whereas linear part is a long range. Coulombic interaction is known from perturbative quantum chromodynamics and the large distance interaction known from lattice QCD [42, 43]. We modify the Cornell potential by adding the harmonic term. The resultant quark-antiquark interaction known as Killingbeck potential which is $ar^2 + br + \dfrac{c}{r}$ [44]. Notice those terms regard as isotonic-type interaction. The energy spectrum of the isotonic potential is isomorphous to the harmonic oscillator spectrum, i.e. it consists of an infinite set of equidistant energy levels. For this reason this oscillator called "the isotonic oscillator". Generalized isotonic oscillators can be seemed as possible representations of realistic quantum dots [39]. The behavior of the Killingbeck plus isotonic oscillator can be seen in Fig. 1.

By substituting Eq. (5) into Eq. (4) we obtain the following equation

$$\frac{d^2 u_{\nu\gamma}(r)}{dr^2} = -2\mu \left[ E - ar^2 - br - \frac{c}{r} - \frac{d}{r^2} - \frac{hr}{r^2+1} - \frac{kr^2}{(r^2+1)^2} - \frac{(2\gamma+5)(2\gamma+3)}{8\mu r^2} \right] u_{\nu\gamma}(r), \qquad (6)$$

And regarding $\dfrac{r^2}{(r^2+1)^2} = \dfrac{1}{r^2+1} - \dfrac{1}{(r^2+1)^2}$ we have

$$\frac{d^2 u_{\nu\gamma}(r)}{dr^2} = [-2\mu E + 2\mu a r^2 + 2\mu b r + 2\mu \frac{c}{r} + 2\mu \frac{d}{r^2} + 2\mu \frac{hr}{r^2+1} + 2\mu \frac{k}{r^2+1}$$
$$-2\mu \frac{k}{(r^2+1)^2} + \frac{(2\gamma+5)(2\gamma+3)}{4r^2}] u_{\nu\gamma}(r), \qquad (7)$$

We suppose the following form for the wave function

$$u_{\nu,\gamma}(r) = g(r)\exp(f(r)). \tag{8}$$

Now for the functions $f(r)$ and $g(r)$ we make use of the ansatz [45-49]

$$g(r) = \begin{cases} 1 & \nu = 0 \\ \prod_i^\nu (r - \alpha_i^\nu) & \nu \geq 1 \end{cases} \tag{9}$$

$$f(r) = \alpha r^2 + \beta r + \lambda \ln r + \eta \ln(r^2+1), \quad \alpha > 0,$$

From Eq. (8) we obtain

$$u''_{\nu,\gamma}(r) = \left[ f''(r) + f'^2(r) + \frac{2f'(r)g'(r) + g''(r)}{g(r)} \right] u_{\nu,\gamma}(r), \tag{10}$$

And from Eq. (9) we have

$$f'(r) = 2\alpha r + \beta + \frac{\lambda}{r} + \frac{2\eta r}{(r^2+1)},$$

$$f'^2(r) = 4\alpha^2 r^2 + \beta^2 + 4\alpha\beta r + \frac{\lambda^2}{r^2} + \frac{4\eta^2 r^2}{(r^2+1)^2} + \frac{4\lambda\eta}{r^2+1} + 4\alpha\lambda + \frac{8\alpha\eta r^2}{r^2+1} + \frac{2\beta\lambda}{r} + \frac{4\beta\eta r}{r^2+1}, \tag{11}$$

$$f''(r) = 2\alpha - \frac{\lambda}{r^2} + 2\eta \frac{(r^2+1) - 2r^2}{(r^2+1)^2},$$

$$f''(r) = 2\alpha - \frac{\lambda}{r^2} + \frac{2\eta}{(r^2+1)} - \frac{4\eta r^2}{(r^2+1)^2}.$$

Regarding $\dfrac{r^2}{(r^2+1)^2} = \dfrac{1}{r^2+1} - \dfrac{1}{(r^2+1)^2}$ and $\dfrac{r^2}{r^2+1} = 1 - \dfrac{1}{r^2+1}$ we have

$$f'^2(r) = 4\alpha^2 r^2 + \beta^2 + 4\alpha\beta r + \frac{\lambda^2}{r^2} + \frac{4\eta^2}{r^2+1} - \frac{4\eta^2}{(r^2+1)^2} + \frac{4\lambda\eta}{r^2+1} + 4\alpha\lambda + 8\alpha\eta - \frac{8\alpha\eta}{r^2+1} + \frac{2\beta\lambda}{r} + 4\beta\eta \frac{r}{r^2+1},$$

$$f''(r) = 2\alpha - \frac{\lambda}{r^2} - \frac{2\eta}{r^2+1} + \frac{4\eta}{(r^2+1)^2}, \tag{12}$$

Substituting of Eqs. (11) and (12) into Eq. (10) leads to

$$u''_{\nu\gamma}(r) = [4\alpha^2 r^2 + \beta^2 + 4\alpha\beta r + \frac{\lambda^2}{r^2} + \frac{4\eta^2}{r^2+1} - \frac{4\eta^2}{(r^2+1)^2} + \frac{4\lambda\eta}{r^2+1}$$

$$+ 4\alpha\lambda + 8\alpha\eta - \frac{8\alpha\eta}{r^2+1} + \frac{2\beta\lambda}{r} + 4\beta\eta \frac{r}{r^2+1} + 2\alpha - \frac{\lambda}{r^2} - \frac{2\eta}{r^2+1} + \frac{4\eta}{(r^2+1)^2}] u_{\nu\gamma}(r), \tag{13}$$

Or

$$u''_{0,\gamma}(r) = [4\alpha^2 r^2 + \beta^2 + 4\alpha\lambda + 8\alpha\eta + 2\alpha + 4\alpha\beta r + \frac{\lambda^2}{r^2} - \frac{\lambda}{r^2} + \frac{2\beta\lambda}{r}$$

$$+ \frac{4\eta^2}{r^2+1} + \frac{4\lambda\eta}{r^2+1} - \frac{2\eta}{r^2+1} - \frac{8\alpha\eta}{r^2+1} - \frac{4\eta^2}{(r^2+1)^2} + \frac{4\eta}{(r^2+1)^2} + 4\beta\eta \frac{r}{r^2+1}] u_{0,\gamma}(r), \tag{14}$$

After some simplicity. By Comparing Eqs. (7) and (14), it can be found that

$$\lambda^2 - \lambda - 2\mu d - \frac{(2\gamma+5)(2\gamma+3)}{4} = 0, \quad 4\alpha^2 = 2\mu a, \quad -4\eta^2 + 4\eta = -2\mu k,$$
$$4\beta\eta = 2\mu h, \quad 4\eta^2 + 4\lambda\eta - 8\alpha\eta - 2\eta = 2\mu k, \quad 4\alpha\beta = 2\mu b, \quad (15)$$
$$2\beta\lambda = 2\mu c, \quad \beta^2 + 4\alpha\lambda + 8\alpha\eta + 2\alpha = -2\mu E,$$

Equation (15) immediately yields

$$\lambda = \frac{1+\sqrt{1+8\mu d + (2\gamma+5)(2\gamma+3)}}{2}, \quad \alpha = -\sqrt{\frac{\mu a}{2}}, \quad \eta = \frac{1+\sqrt{1+2\mu k}}{2},$$
$$\beta = \frac{\mu h}{2\eta}, \quad a = \frac{(2\lambda+1)^2}{8\mu}, \quad b = \frac{2\alpha\beta}{\mu}, \quad c = \frac{\beta\lambda}{\mu}, \quad (16)$$

And the energy can be obtained by

$$E_{\nu\gamma} = -\frac{1}{2\mu}(\beta^2 + 4\alpha\lambda + 8\alpha\eta + 2\alpha), \quad (17)$$

The spin and isospin dependent interactions are not the only source of *SU* (6) violation. In order to study the baryon spectrum one has to consider the *SU* (3) violation produced by the differences in the quark masses. The Gell-Mann-Okubo (GMO) mass formula [50] made use of a λ$_8$ violation of *SU* (3) in order to explain the mass splitting within the various *SU* (3) multiplets. The hypercentral constituent quark model is fairly good for description the baryon spectrum [51], but in some cases the splitting within the various *SU* (6) multiplets are too low. The preceding results [32, 52, 53] show that both spin and isospin dependent terms in the quark Hamiltonian are important. Description of the splitting within the *SU* (6) baryon multiplets are presented by the Gürsey Radicati mass formula [33]:

$$M = M_0 + CC_2[SU_S(2)] + DC_1[U_Y(1)] + I[C_2[SU_I(2)] - \frac{1}{4}(C_1[U_Y(1)])^2], \quad (18)$$

Where $M_0$ is the average energy value of the *SU* (6) multiplet, $C_2[SU_S(2)]$ and $C_2[SU_I(2)]$ are the *SU* (2) Casimir operators for spin and isospin, respectively, and $C_1[U_Y(1)]$ is the Casimir operator for the *U* (1) subgroup generated by the hypercharge *Y* [54, 55]. This mass formula has tested to be successful in the description of the ground state baryon masses, however, as stated by the authors themselves, it is not the most general mass formula that can be written on the basis of a broken *SU* (6) symmetry. In order to generalize Eq. (18), Giannini and et al considered dynamical spin- flavor symmetry $SU_{SF}$ (6) [34] and described the $SU_{SF}$ (6) symmetry breaking mechanism by generalizing Eq. (18) as:

$$M = M_0 + AC_2[SU_{SF}(6)] + BC_2[SU_F(3)] + CC_2[SU_S(2)] + DC_1[U_Y(1)]$$
$$+ I[C_2[SU_I(2)] - \frac{1}{4}(C_1[U_Y(1)])^2], \quad (19)$$

In Eq. (19) the spin term $(CC_2[SU_S(2)])$ represents the spin-spin interactions, the flavor term $(BC_2[SU_F(3)])$ denotes the flavor dependence of the interactions, and the $SU_{SF}$ (6) term $(AC_2[SU_{SF}(6)])$ depends on the permutation symmetry of the wave functions, represents "signature - dependent" interactions [54]. The last two terms $\left(I[C_2[SU_I(2)] - \frac{1}{4}(C_1[U_Y(1)])^2]\right)$ represent the isospin and hypercharge dependence of the masses. In Table 1, we give the expectation values of the Casimir operators $SU_{SF}(6)$ and $SU_F(3)$ for the allowed three-quark configurations.

The generalized Gürsey Radicati mass formula Eq. (19) can be used to describe the octet and decuplet baryons spectrum, provided that two conditions are fulfilled. The first condition is the feasibility of using the same splitting coefficients for different $SU(6)$ multiplets. This seems actually to be the case, as shown by the algebraic approach to the baryon spectrum [1]. The second condition is given by the feasibility of getting reliable values for the unperturbed mass values $M_0$ [29]. For this purpose we regarded the $SU(6)$ invariant part of the hCQM, which provides a good description of the baryon spectrums and used the Gürsey Radicati inspired $SU(6)$ breaking interaction to describe the splitting within each $SU(6)$ multiplet. Therefore, the baryons masses are obtained by three quark masses and the Eigen energies ($E_{\nu\gamma}$) of the radial Schrödinger equation with the expectation values of $H_{GR}$ as follows:

$$M = 3m + E_{\nu\gamma} + \langle H_{GR} \rangle. \tag{20}$$

In the above equation $m$ is the reduced mass. The $H_{GR}$ is in the following form:

$$H_{GR} = AC_2[SU_{SF}(6)] + BC_2[SU_F(3)] + CC_2[SU_S(2)] + DC_1[U_Y(1)] \\ + I\left[C_2[SU_I(2)] - \frac{1}{4}(C_1[U_Y(1)])^2\right]. \tag{21}$$

The expectation values of $H_{GR}$ ($\langle H_{GR} \rangle$), is completely identified by the expectation values of the Casimir operators (Table 1).
Therefor we have:

$$M = 3m + E_{\nu\gamma} + A\langle C_2[SU_{SF}(6)]\rangle + B\langle C_2[SU_F(3)]\rangle + C\langle C_2[SU_S(2)]\rangle + D\langle C_1[U_Y(1)]\rangle \\ + I\left[\langle C_2[SU_I(2)]\rangle - \frac{1}{4}\langle (C_1[U_Y(1)])\rangle^2\right]. \tag{22}$$

In order to simplify the solving procedure, the constituent quarks masses are assumed to be the same for Up, Down and Strange quark flavors ($m = m_u = m_d = m_s$), therefore, within this approximation, the $SU(6)$ symmetry is only broken dynamically by the spin and flavour dependent terms in the Hamiltonian. We determined $E_{\nu\gamma}$ by exact solution of the radial Schrödinger equation for the hypercentral Potential Eq. (5). For calculating the baryons mass according to Eq. (22), we need to find the unknown parameters. For this purpose we choose a limited number of well-known resonances and express their mass differences using $H_{GR}$ and the Casimir operator expectation values: $N(1650)S11 - N(1535)S11 = 3C$, $4N(938)P11 - \Sigma(1193)P11 - 3\Lambda(1116)P01 = 4D$ and $\Sigma(1193)P11 - \Lambda(1116)P01 = 2I$. Leading to the numerical values: $C = 38.3$, $D = -197.3$ $MeV$ and $I = 38.5$ MeV. For determining $m, \alpha, \beta, d$ and $\eta$ ( in Eq. (17) ) and the two coefficients $A$ and $B$ of Eq. (19) we have used the Newton-Raphson Method for solving the nonlinear equations. For our purpose we chose $N(938)P11$, $\Delta(1232)P33$, $\Lambda(1116)P01$, $\Sigma(1193)P11$, $\Lambda(1810)P01$, $\Delta(1700)D33$ and $\Sigma(1940)D13$ which yielded the best reproduction (the maximum percentage of relative error is 0.33 %), then by solving seven nonlinear equations with seven unknown parameters we calculated the free parameters ($m, \alpha, \beta, d, \eta, A, B$). The fitted parameters are reported in Table 2. The corresponding numerical values for 3 and 4 star baryons resonances are given in Tables 3 and 4, column $M_{ourCalc}$.

In Tables 3 and 4, column $M_{[34]Calc}$, we've shown the numerical values of the calculated masses of baryon resonances by Giannini and et al, where they regarded the confinement potential as the Cornell potential ($-\frac{\tau}{x} + \alpha x$). The solution of the hypercentral Schrödinger equation with this potential cannot be obtained analytically [34], therefore Giannini and et al used the dynamic symmetry $O(7)$ of the hyperCoulomb problem to obtain the hyperCoulomb Hamiltonian and Eigen functions analytically and also they regarded the linear term as a perturbation. Comparison between our results and the experimental masses [51] show that our model has improved the results of model in Ref. [34], particularly in $\Lambda(1810)$, $\Lambda(2110)$ F05, $\Lambda^*(1405)$ S01, $\Lambda^*(1520)$ D01, $\Delta(1905)$ F35, $\Delta(1910)$ P31, $\Delta(1920)$ P33 and $\Sigma(1775)$ D15 (refer to Tables 3 and 4). These improvements in reproduction of baryons resonance masses obtained by using a suitable form for confinement potential and exact analytical solution of the radial Schrödinger equation for our proposed potential. The percentage of relative error for our calculations is between 0 and 10 % (column 7, in Table 3 and 4). The corresponding numerical values for some of 1 and 2 star baryons resonances mass up to 2.1 GeV are given in Table 5, column $M_{ourCalc}$. The percentage of relative error for our calculations is between 0.07 and 9 % (column 6, in Table 5). Comparison between our results and the experimental masses [56] show that the baryon spectrums are, in general, fairly well reproduced.

## 3. CONCLUSION

In this paper we have computed the baryon resonances spectrum up to 3 GeV within a non-relativistic quark model based on the three identical quarks Schrödinger equation and the algebraic approach. We have solved the Schrodinger equation numerically to obtain the energy Eigen values under the Killingbeck plus isotonic oscillator interaction potentials. Then, we fitted the generalized GR mass formula parameters to the baryons energies and calculated the baryon masses. The overall good description of the spectrum which we obtain by our proposed model shows that our theoretical model can also be used to give a fair description of the energies of the excited multiplets up to 3 GeV and not only for the ground state octets but also decuplets. Moreover, our model reproduces the position of the Roper resonances of the nucleon and negative-parity resonance. There are problems in the reproduction of the experimental masses in $\Delta(1620)$ S31 and $\Sigma(1670)$ D13 turn out to have predicted mass about 100 MeV above the experimental value. A better agreement may be obtained either using the square of the mass [1] or trying to include a spatial dependence in the *SU*(6)-breaking part.

## REFERENCES

1. R. Bijker, F. Iachello and A. Leviatan, Ann. Phys. (N.Y.) 236, 69 (1994).
2. E. Santopinto and M.M. Giannini, Phys. Rev. C86, 065202 (2012).
3. B. Chakrabarti, A. Bhattacharya, S. Mani, A. Sagari, Acta Physica Polonica B, 41, 95-101(2010).
4. S. Aoki et al., Phys. Lett. 84, 238 (2000).
5. L. I. Abou-Salem, Advances in High Energy Physics, Vol 2014, Article ID 196484, 1-5 (2014).
6. N. Salehi, H. Hassanabadi, and A.A. Rajabi, *Eur. Phys. J. Plus* **128**, 27 (2013).
7. N. Salehi, H. Hassanabadi, and A.A. Rajabi, *Chinese Physics C* **37**, 113101 (2013).
8. M. M. Giannini and E. Santopinto, *Chin. J. Phys.* **53**, 020301(2015).

9. 6. Gell-Mann, M., and Y. Ne'eman, 1964, The eightfold way,Frontiers in Physics (Benjamin, New York, NY).
10. Kokkedee, J. J. J., 1969, The quark model, Frontiers inPhysics (Benjamin, New York, NY), based on a series oflectures given at CERN in Autumn 1967.
11. Narison, S., 2004, QCD as a theory of hadrons: from partonsto confinement; electronic version (Cambridge Univ. Press,Cambridge).
12. Yndur´ain, F. J., 1999, The theory of quark and gluon interactions; 3rd ed., Texts and monographs in physics (Springer,Berlin).
13. A.M.Abazov et al ., Phys. Rev. Lett. 99 (2007) 052001
14. T. Aaltonen et al ., Phys. Rev. Lett. 99 (2007) 052002.
15. K.C. Bowler et al . (UKQCD Collaboration), Phys. Rev. D 54 (1996) 3619.
16. R. Lewis et al ., Phys. Rev. D 64 (2001) 094509, Phys. Rev. D 64 (2001) 094509.
17. N. Mathur et al ., Phys. Rev. D 66 (2002) 014502.
18. S. Gottlieb and S. Tamhankar, Nucl. Phys. Proc. Suppl. 119 (2003) 644.
19. A. Ali Khan et al ., Phys. Rev. D 62 (2000) 054505.
20. J. M. Flynn et al ., JHEP 7 (2003) 66.
21. H. Na and S. Gottlieb, PoS(LAT2006)191 [hep-lat/0610009].
22. H. Na and S. Gottlieb, PoS(LAT2007)124, arXiv:0710.1422 [hep-lat]
23. R. Lewis and R. M. Woloshyn, arXiv:0806.4783 [hep-lat]
24. D. M. Asneret al., Int. J. Mod. Phys. A 24,1 (2009).
25. G. Wilson, Phys. Rev. D10, 2445 (1974).
26. S. Weinberg, Trans. N.Y. Acad. Sci.38, 185 (1977).
27. S. Aoki, Nucl. Phys., Sect. B Proc.Suppl.94, 3 (2001); D. Toussaint, Nucl. Phys., Sect. BProc. Suppl.106, 111 (2002); T. Kaneko, Nucl. Phys.,Sect. B Proc. Suppl.106, 133 (2002).
28. CP-PACS Collaboration, S. Aoki et al., Phys. Rev. Lett.84, 238 (2000); Phys. Rev. D 67, 034503(2003).
29. Gunnar S. Bali et al., Phys. Rev. D62, 054503 (2000).
30. C. Alexandrou, P. de Forcrand and O. Jahn, *Nucl. Phys. Proc.* Suppl. **119**, 667 (2003).
31. E. Santopinto, F. Iachello and M. M. Giannini, *Eur. Phys. J.* **A 1**, 307 (1998).
32. M. M. Giannini, E. Santopinto and A, Vassallo, *Eur. Phys. J.* **A12**, 447 (2001).
33. F. Gürsey and L. A. Radicati , *Phys. Rev. Lett.* **13**, 173 (1964).
34. M. M. Giannini, E. Santopinto, and A.Vassallo, *Eur.Phys.J.* **A25,** 241-247(2005).
35. N. Salehi, A.A. Rajabi, *Modern Physics Letters A*, Vol. 24, No. **32**, 2631-2637 (2009).
36. N. Salehi, A.A. Rajabi, *Phys. Scr.* **85**, 055101 (2012).
37. N. Saad, R. L. Hall, H. Ciftci and O. Yesiltas, Adv. Math. Phys. 750168, (2011).
38. R. L. Hall, N. Saad and O. Yesiltas, J. Phys. A: Math. Theor. 43, 465304, (2010).
39. D. Agboola, J. Links, I. Marquette and Y. Z. Zhang, J. Phys. A: Math. Theor. 47, 395305, (2014).
40. N. Salehi, H. Hassanabadi, A.A. Rajabi, *Eur. Phys. J. Plus*, 128:27(2013).
41. M. V. N. Murthy, Z. *Phys.* **C31**, 81 (1986).
42. K. Zalewski, *Acta Phys.Polon. B*, 29 2535-2538 (1998).
43. Y-B. Ding, X-Q. Li, P-N. Shen, Phys. Rev. D 60: 074010 (1999)M. Znojil, *J. Math Phys.* **31**, (1990).
44. S. Zaim et al., arXiv: 1410.0399v1 (2014).
45. M. Znojil, J. Math Phys. 31, (1990).
46. A. A. Rajabi, *Indian J. Pure and Appl. Phys.* **l41**, 89 (2003).
47. A. A. Rajabi and N. Salehi, *Iranian Journal of Physics Research*, **8(3)**, 169-175 (2008).
48. N. Salehi, H. Hassanabadi, *Romanian Reports in Physics*, Vol. 67, No. 2, (2015).
49. N. Salehi, H. Hassanabadi, *Int. J. Mod. Phys. E,* **24**, (2015).
50. See e.g. M. Gell-Mann and Y. Ne'eman, The eightfold way (W.A. Benjamin, Inc., New York, 1964).
51. M. Ferraris et al., Phys. Lett. B**364**, 231 (1995).
52. N. Salehi, A.A. Rajabi, *Mod. Phys. Lett. A,* **24**, 2631 (2009).
53. H. Hassanabadi, A. A. Rajabi, *Modern Physics Letters A,* Vol. **24**, Nos. 11-13, 1043-1046 (2009).
54. F. J. Squires, Nuovo Cimento 25 (1962), 242; *A.Martin, Phys. Lett. 1*, **72** (1962).
55. R. Bijker, M. M. Giannini and E. Santopinto, *Eur.Phys. J.* A **22**, 319 (2004).
56. K.A. Olive et al. (Particle Data Group), *Chin. Phys.* C **38**, 090001 (2014).

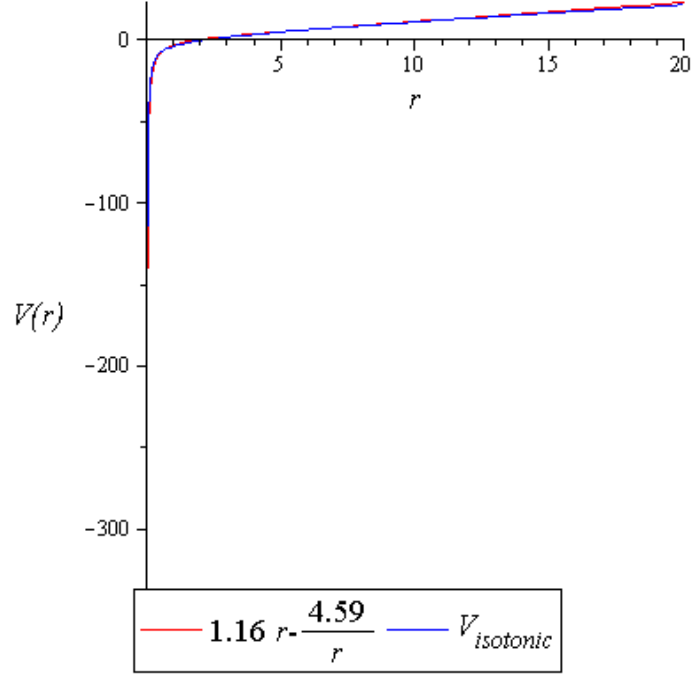

Figure 1. V(r) (in MeV) with a=0.0072, b=1.239, c=-3.8149, d=0.0032, h= -2.7904 and k=1.01008.

Table 1. Eigenvalues of the $C_2[SU_{SF}(6)]$ and $C_2[SU_F(3)]$ Casimir operators.

| *Dimension* ( *SU*(6) ) | $C_2[SU_{SF}(6)]$ | *Dimension*( *SU*(3) ) | $C_2[SU_F(3)]$ |
|---|---|---|---|
| 56 | $\frac{45}{4}$ | 8 | 3 |
| 70 | $\frac{33}{4}$ | 10 | 6 |
| 20 | $\frac{21}{4}$ | 1 | 0 |

Table 2. The fitted values of the parameters of the Eq. (22) for N, $\Delta$, $\Lambda, \Sigma$, $\Xi$ and $\Omega$ baryons, obtained with resonances mass differences and global fit to the experimental resonance masses [56].

| **Parameter** | A | B | C | D | I | m | $\alpha$ | $\beta$ | d | $\eta$ |
|---|---|---|---|---|---|---|---|---|---|---|
| **Value** | -19.616 MeV | 18.575 *MeV* | 38.3 | -197.3 *MeV* | 38.5 *MeV* | 271 *MeV* | -0.381 $MeV^2$ | 0.489 *MeV* | 0.388 | 0.446 |

10Table 3. Mass spectrum of baryons resonances (in MeV) calculated with the mass formula Eq. (22). The column $M_{Our\,Calc}$ contains our calculations with the parameters of table 2 and the column $M_{[34]Calc}$ show calculations of Giannini et al. The column 7 indicates the percentage of relative error for our calculations.

| Baryon | Status | $M_{[56]exp}$ | State | $M_{[34]Calc}$ | $M_{Our\,Calc}$ | Percent of relative error for Our Calculation |
|---|---|---|---|---|---|---|
| N(938) P11 | **** | 938 | $^2 8_{1/2}[56, 0^+]$ | 938 | 938 | 0% |
| N(1440) P11 | **** | 1410-1450 | $^2 8_{1/2}[56, 0^+]$ | 1448.7 | 1448.51 | 2.73% - 0.1% |
| N(1520) D13 | **** | 1510-1520 | $^2 8_{3/2}[70, 1^-]$ | 1543.7 | 1528.79 | 1.24% - 0.57% |
| N(1535) S11 | **** | 1525-1545 | $^2 8_{1/2}[70, 1^-]$ | 1543.7 | 1528.79 | 0.24% - 1.04% |
| N(1650) S11 | **** | 1645-1670 | $^4 8_{1/2}[70, 1^-]$ | 1658.6 | 1643.69 | 0.07% - 1.57% |
| N(1675) D15 | **** | 1670-1680 | $^4 8_{5/2}[70, 1^-]$ | 1658.6 | 1643.69 | 1.57% - 2.16% |
| N(1680) F15 | *** | 1680-1690 | $^2 8_{5/2}[56, 2^+]$ | 1651.4 | 1688.57 | 0.51% - 0.08% |
| N(1700) D13 | *** | 1650-1750 | $^4 8_{3/2}[70, 1^-]$ | 1658.6 | 1643.69 | 0.38% - 6.07% |
| N(1710) P11 | *** | 1680-1740 | $^2 8_{1/2}[56, 0^+]$ | 1795.4 | 1798.16 | 7.03% - 3.34% |
| N(1720) P13 | **** | 1700-1750 | $^2 8_{3/2}[56, 2^+]$ | 1651.4 | 1688.57 | 0.67% - 3.51% |
| N(1875) D13 | *** | 1820-1920 | $^2 8_{3/2}[70, 1^-]$ | … | 1857.01 | 2.03% - 3.28% |
| N(1900) P13 | *** | 1875-1935 | $^2 8_{3/2}[70, 2^+]$ | … | 1966.7 | 4.89% - 1.63% |
| N(2190) G17 | **** | 2100-2200 | $^2 8_{7/2}[70, 3^-]$ | … | 2186.37 | 4.11% - 0.61% |
| N(2220) H19 | **** | 2200-2300 | $^2 8_{9/2}[56, 4^+]$ | … | 2237.44 | 1.7% - 2.72% |
| N(2250) G19 | **** | 2200-2350 | $^4 8_{9/2}[70, 3^-]$ | … | 2301.27 | 4.6% - 2.07% |
| N(2600) I1,11 | *** | 2550-2750 | $^2 8_{11/2}[70, 5^-]$ | … | 2626.25 | 2.99% - 4.5% |
| Δ (1232) P33 | **** | 1230-1234 | $^4 10_{3/2}[56, 0^+]$ | 1232 | 1232.37 | 0.19% - 0.13% |
| Δ (1600) P33 | *** | 1500-1700 | $^4 10_{3/2}[56, 0^+]$ | 1683 | 1647.2 | 9.81% - 3.1% |
| Δ (1620) S31 | **** | 1600-1660 | $^2 10_{1/2}[70, 1^-]$ | 1722.8 | 1700.01 | 6.25% - 2.41% |
| Δ (1700) D33 | **** | 1670-1750 | $^2 10_{3/2}[70, 1^-]$ | 1722.8 | 1700.01 | 1.79% - 2.85% |
| Δ (1905) F35 | **** | 1855-1910 | $^4 10_{5/2}[56, 2^+]$ | 1945.4 | 1865.27 | 0.55% - 2.34% |
| Δ (1910) P31 | **** | 1860-1910 | $^4 10_{1/2}[56, 2^+]$ | 1945.4 | 1865.27 | 0.28% - 2.34% |
| Δ (1920) P33 | *** | 1900-1970 | $^4 10_{3/2}[56, 0^+]$ | 2089.4 | 1974.7 | 3.93% - 0.23% |
| Δ (1930) D35 | *** | 1900-2000 | $^2 10_{5/2}[70, 2^-]$ | … | 1918.6 | 0.97% - 4.07% |
| Δ (1950) D35 | **** | 1915-1950 | $^4 10_{7/2}[56, 2^+]$ | 1945.4 | 1865.27 | 2.59% - 4.34% |
| Δ (2420) H3, 11 | **** | 2300-2500 | $^4 10_{11/2}[56, 4^+]$ | … | 2303.79 | 0.16% - 7.84% |

Table 4. As Table 3, but for $\Lambda$, $\Sigma$, $\Xi$ and $\Omega$ resonances.

| Baryon | Status | $M_{[56]exp}$ | State | $M_{[34]Calc}$ | $M_{Our\,Calc}$ | Percent of relative error for Our Calculation |
|---|---|---|---|---|---|---|
| $\Lambda$ (1116)P01 | **** | 1116 | $^2 8_{1/2}[56, 0^+]$ | 1116 | 1116.05 | 0.004% |
| $\Lambda$ (1600)P01 | *** | 1560-1700 | $^2 8_{1/2}[56, 0^+]$ | 1626.7 | 1647.99 | 5.64% - 3.05% |
| $\Lambda$ (1670)S01 | **** | 1660-1680 | $^2 8_{1/2}[70, 1^-]$ | 1721.7 | 1706.84 | 2.82% - 1.59% |
| $\Lambda$ (1690)D03 | **** | 1685-1695 | $^2 8_{3/2}[70, 1^-]$ | 1721.7 | 1706.84 | 1.29% - 0.69% |
| $\Lambda$ (1800)S01 | *** | 1720-1850 | $^4 8_{1/2}[70, 1^-]$ | 1836.6 | 1821.74 | 5.91% - 1.52% |
| $\Lambda$ (1810)P01 | *** | 1750-1850 | $^2 8_{1/2}[70, 0^+]$ | 1973.4 | 1816.04 | 3.77% - 1.83% |
| $\Lambda$ (1820) F05 | **** | 1815-1825 | $^2 8_{5/2}[56, 2^+]$ | 1829.4 | 1866.62 | 2.84% - 2.28% |
| $\Lambda$ (1830)D05 | **** | 1810-1830 | $^4 8_{5/2}[70, 1^-]$ | 1836.6 | 1821.74 | 0.64% - 0.45% |
| $\Lambda$ (1890)P03 | **** | 1850-1910 | $^2 8_{3/2}[56, 2^+]$ | 1829.4 | 1866.62 | 0.89% - 2.27% |
| $\Lambda$ (2100)G07 | **** | 2090-2110 | $^2 1_{7/2}[70, 3^-]$ | … | 2089.04 | 0.04% - 0.99% |
| $\Lambda$ (2110)F05 | **** | 2090-2140 | $^4 8_{5/2}[70, 2^+]$ | 1995 | 2149.96 | 2.86% - 0.46% |
| $\Lambda$ (2350) H09 | *** | 2340-2370 | $^2 8_{9/2}[56, 4^+]$ | … | 2360.52 | 0.87% - 0.4% |
| $\Lambda^*$(1405) S01 | **** | 1402-1410 | $^2 1_{1/2}[70, 1^-]$ | 1657.9 | 1433.91 | 2.27% - 1.69% |
| $\Lambda^*$(1520)D01 | **** | 1518-1520 | $^2 1_{3/2}[70,1^-]$ | 1657.9 | 1433.91 | 5.53% - 5.66% |
| $\Sigma$ (1193) P11 | **** | 1193 | $^2 8_{1/2}[56, 0^+]$ | 1193 | 1193.05 | 0.004% |
| $\Sigma$ (1660)P11 | *** | 1630-1690 | $^2 8_{1/2}[56, 0^+]$ | 1703.7 | 1616.12 | 0.05% - 4.37% |
| $\Sigma$ (1670)D13 | **** | 1665-1685 | $^2 8_{3/2}[70, 1^-]$ | 1798.7 | 1783.74 | 7.13% - 5.85% |
| $\Sigma$ (1750)S11 | *** | 1730-1800 | $^2 8_{1/2}[70, 1^-]$ | 1798.7 | 1783.74 | 3.1% - 0.9% |
| $\Sigma$ (1775) D15 | **** | 1770-1780 | $^4 8_{5/2}[70, 1^-]$ | 1913.6 | 1789.87 | 1.12% - 0.55% |
| $\Sigma$ (1915)F15 | **** | 1900-1935 | $^2 8_{5/2}[56, 2^+]$ | 1906.4 | 1910.7 | 0.56% - 1.25% |
| $\Sigma$ (1940)D13 | *** | 1900-1950 | $^2 8_{3/2}[56, 1^-]$ | 1913.6 | 1943.62 | 2.29% - 0.32% |
| $\Sigma^*$(1385)P13 | **** | 1383-1385 | $^4 10_{3/2}[56, 0^+]$ | … | 1363.67 | 1.39% - 1.54% |
| $\Sigma^*$(2030)F17 | **** | 2025-2040 | $^4 10_{7/2}[56, 2^+]$ | 2085.0 | 2004.82 | 0.99% - 1.72% |
| $\Xi$ (1318) P11 | **** | 1314-1316 | $^2 8_{1/2}[56, 0^+]$ | 1332.6 | 1332.6 | 1.41% - 1.26% |
| $\Xi$ (1690) S11 | *** | 1680-1700 | $^2 8_{1/2}[70, 1^-]$ | … | 1706.1 | 1.55% - 0.35% |
| $\Xi$ (1820) D13 | *** | 1818-1828 | $^2 8_{3/2}[70, 1^-]$ | 1938.3 | 1923.39 | 5.79% - 5.21% |
| $\Xi^*$(1530) P13 | **** | 1531-1532 | $^4 10_{3/2}[56, 0^+]$ | 1511.1 | 1503.2 | 1.81% - 1.87% |
| $\Omega$ (1672) P03 | **** | 1672-1673 | $^4 10_{3/2}[56, 0^+]$ | 1650.7 | 1643 | 1.73% - 1.79% |
| $\Omega$ (2250) D03 | *** | 2243-2261 | $^2 10_{3/2}[70, 1^-]$ | … | 2227.87 | 0.67% - 1.46% |

Table 5. Mass spectrum of some of 1 and 2 star baryons resonances (in MeV) up to 2.1 GeV calculated with the mass formula Eq. (22). The column $M_{Our\,Calc}$ contains our calculations with the parameters of Table 2 and the 6 indicates the percentage of relative error for our calculations.

| Baryon | Status | $M_{[56]exp}$ | State | $M_{Our\,Calc}$ | Percent of relative error for Our Calculation |
|---|---|---|---|---|---|
| N(1860)F15 | ** | 1820-1960 | $^2 8_{5/2}[70, 2^+]$ | 1966.7 | 8.06% - 0.34% |
| N(1880)P11 | ** | 1835-1905 | $^4 8_{1/2}[70, 2^+]$ | 1971.9 | 7.46% -3.51 % |
| N(1895)S11 | ** | 1880-1910 | $^2 8_{1/2}[70, 1^-]$ | 1857.01 | 1.22% - 2.77% |
| N(1990)F17 | ** | 1995-2125 | $^4 8_{7/2}[70, 2^+]$ | 1971.9 | 1.15% - 7.2% |
| N(2000)F15 | ** | 1950-2150 | $^4 8_{5/2}[70, 2^+]$ | 1971.9 | 1.12% - 8.28% |
| N(2040)P13 | * | 2031-2065 | $^4 8_{3/2}[70, 2^+]$ | 1971.9 | 2.9% - 4.5% |
| N(2060)D15 | ** | 2045-2075 | $^4 8_{5/2}[70, 2^-]$ | 1971.9 | 3.57% - 4.96% |
| N(2120)D13 | ** | 2090-2210 | $^2 8_{3/2}[56, 1^-]$ | 2127.52 | 1.79% - 3.73% |
| Δ(1750)P31 | * | 1708-1780 | $^2 10_{1/2}[70, 0^+]$ | 1754.5 | 2.72% - 1.43% |
| Δ(1900)S31 | ** | 1840-1920 | $^2 10_{1/2}[70, 1^-]$ | 1918.6 | 4.27% - 0.07% |
| Δ(1940)D33 | ** | 1940-2060 | $^2 10_{3/2}[70, 1^-]$ | 1918.6 | 1.1% - 6.86% |
| Δ(2000)F35 | ** | ≈ 2000 | $^2 10_{5/2}[70, 2^+]$ | 2028.2 | 1.41% |
| Σ(1580)D13 | * | ≈ 1580 | $^4 8_{3/2}[70, 1^-]$ | 1574.12 | 0.37% |
| Σ(1620)S11 | ** | ≈ 1620 | $^2 8_{1/2}[70, 0^-]$ | 1674.97 | 3.39% |
| Σ(1770)P11 | * | ≈ 1770 | $^2 8_{1/2}[70, 0^+]$ | 1783.84 | 0.78% |
| Σ(1880)P11 | ** | ≈ 1880 | $^2 8_{1/2}[20, 1^+]$ | 1842.68 | 1.98% |
| Σ(2000)S11 | * | ≈ 2000 | $^2 8_{1/2}[70, 1^-]$ | 2002.47 | 0.12% |
| Σ(2070)F15 | * | ≈ 2070 | $^4 8_{5/2}[70, 2^+]$ | 2117.37 | 2.28% |
| Σ$^*$(1840)P13 | * | ≈ 1840 | $^4 10_{3/2}[56, 0^+]$ | 1895.6 | 3.02% |
| Σ$^*$(2080)P13 | ** | ≈ 2080 | $^2 10_{3/2}[70, 2^+]$ | 2058.2 | 1.04% |